\documentclass{article} 
\usepackage{nips10submit_e,times}
\usepackage{amssymb}
\usepackage{amsmath,amsfonts}

\title{Closed-Form Solutions to A Category of Nuclear Norm Minimization Problems}

\author{
Guangcan Liu $^\dag$, Ju Sun $^\ddag$ and Shuicheng Yan $^\S$\\
Department of Electrical \& Computer Engineering, National University of Singapore\\
$^\dag$ \texttt{roth@sjtu.edu.cn}, $^\ddag$ \texttt{idmsj@nus.edu.sg}, $^\S$ \texttt{eleyans@nus.edu.sg} }

\newtheorem{theorem}{Theorem}[section]
\newtheorem{lemma}{Lemma}[section]

\newtheorem{corollary}{Corollary}[section]

\newcommand{\Id}{\mathtt{I}}

\newcommand{\trace}[1]{\ensuremath{\mathrm{tr}\left(#1\right)}}

\newcommand{\spn}[1]{\ensuremath{\mathrm{span}\left(#1\right)}}
\newcommand{\rank}[1]{\ensuremath{\mathrm{rank}\left(#1\right)}}
\newcommand{\norm}[1]{\ensuremath{\left\| #1 \right\|}}

\newcommand{\qed}{\nobreak \ifvmode \relax \else
      \ifdim\lastskip<1.5em \hskip-\lastskip
      \hskip1.5em plus0em minus0.5em \fi \nobreak
      \vrule height0.5em width0.5em depth0.1em\fi}

\newenvironment{proof}[1][Proof]{\begin{trivlist}
\item[\hskip \labelsep {\bfseries #1}]}{\end{trivlist}}
\nipsfinalcopy 

\begin{document}

\maketitle
\begin{abstract}\footnote{The content of this paper is a part of \cite{lrr_extention}} It is an efficient and effective strategy to utilize the nuclear norm approximation to learn low-rank
matrices, which arise frequently in machine learning and computer vision. So the exploration of nuclear norm
minimization problems is gaining much attention recently. In this paper we shall prove that the following Low-Rank
Representation (LRR) \cite{icml_2010_lrr,lrr_extention} problem:
\begin{eqnarray*}
\min_{Z} \norm{Z}_*, & {s.t.,} & X=AZ,
\end{eqnarray*}
has a unique and closed-form solution, where $X$ and $A$ are given matrices. The proof is based on proving a lemma that
allows us to get closed-form solutions to a category of nuclear norm minimization problems.
\end{abstract}

\section{Introduction}
In real applications, our observations are often noisy, or even grossly corrupted, and some observations may be
missing. This fact naturally leads to the problem of recovering a low-rank matrix $X^0$ from a corrupted observation
matrix $X=X^0+E^0$ (each column of $X$ is an observation vector), with $E^0$ being the unknown noise. Due to the
low-rank property of $X^0$, it is straightforward to consider the following regularized rank minimization problem:
\begin{eqnarray*}
\min_{D,E} \hspace{0.1in} \rank{D}+\lambda\norm{E}_{\ell},& \textrm{s.t.} & X=D+E,
\end{eqnarray*}
where $\lambda>0$ is a parameter and $\norm{\cdot}_{\ell}$ is some kind of regularization strategy, such as the
$\ell_1$-norm adopted by \cite{nips_2009_rpca,journal_2009_rpca2}, for characterizing the noise $E^0$. As a common
practice in rank minimization problems, one could replace the rank function with the nuclear norm, resulting in the
following convex optimization problem:
\begin{eqnarray}\label{eq:rpca}
\min_{D,E}\hspace{0.1in} \norm{D}_*+\lambda\norm{E}_{\ell}, &\textrm{s.t.}& X=D+E.
\end{eqnarray}
The minimizer $D^*$ (with respect to the variable $D$) gives a low-rank recovery to the original data $X^0$. The above
formulation is adopted by the recently established Robust PCA (RPCA) method \cite{nips_2009_rpca,journal_2009_rpca2},
which uses the $\ell_1$-norm to characterize the noise. However, such a formulation implicitly assumes that the
underlying data structure is a single low-rank subspace. When the data is drawn from a union of multiple subspaces,
denoted as $\mathcal{S}_1,\mathcal{S}_2,\cdots,\mathcal{S}_k$, the PRCA method actually treats the data as being
sampled from a single subspace defined by $\mathcal{S}=\sum_{i=1}^{k}\mathcal{S}_i$. The specifics of the individual
subspaces are not well considered, so the recovery may be inaccurate.

To better handle the mixed data, in \cite{icml_2010_lrr,lrr_extention} we suggest a more general rank minimization
problem defined as follows:
\begin{eqnarray*}
\min_{Z,E}\hspace{0.1in} \rank{Z}+\lambda\norm{E}_{\ell},& \textrm{s.t.}& X=AZ+E,
\end{eqnarray*}
where $A$ is a ``dictionary'' that linearly spans the data space. By replacing the rank function with the nuclear norm,
we have the following convex optimization problem:
\begin{eqnarray}\label{eq:general:lrr}
\min_{Z,E} \hspace{0.1in} \norm{Z}_*+\lambda\norm{E}_{\ell}, & \textrm{s.t.}& X=AZ+E.
\end{eqnarray}
After obtaining an optimal solution $(Z^*,E^*)$, we could recover the original data by using $AZ^*$ (or $X-E^*$). Since
$\rank{AZ^*}\leq\rank{Z^*}$, $AZ^*$ is also a low-rank recovery to the original data $X^0$. By setting $A=\Id$, the
formulation \eqref{eq:general:lrr} falls backs to \eqref{eq:rpca}. So our LRR method could be regarded as a
generalization of RPCA that essentially uses the standard basis as the dictionary. By choosing an appropriate
dictionary $A$, as shown in \cite{icml_2010_lrr,lrr_extention}, the lowest-rank representation also reveals the
segmentation of data such that LRR could handle well the data drawn from a mixture of multiple subspaces.

For ease of understanding the LRR method, in this work we consider the ``ideal'' case that the data is noiseless. That
is, we consider the following optimization problem:
\begin{eqnarray}\label{eq:lr:nuclear_norm_minization}
\min_{Z} \hspace{0.1in}  \norm{Z}_*, & \textrm{s.t.} & X=AZ.
\end{eqnarray}
We will show that this optimization problem \emph{always} has a unique and closed-form minimizer, provided that $X=AZ$
has feasible solutions.
\section{A Closed-Form Solution to Problem \eqref{eq:lr:nuclear_norm_minization}}
The nuclear norm is convex, but not strongly convex. So it is possible that problem
\eqref{eq:lr:nuclear_norm_minization} has multiple optimal solutions. Fortunately, it can be proven that the minimizer
to problem \eqref{eq:lr:nuclear_norm_minization} is \emph{always} uniquely defined by a closed form. This is summarized
in the following theorem.
\begin{theorem}[Uniqueness]\label{theorem:unique:nonoise}
Assume $A\neq0$ and $X=AZ$ have feasible solutions, i.e., $X\in\spn{A}$. Then
\begin{eqnarray}\label{eq:lrr:close_form}
Z^*=V_A(V_A^TV_A)^{-1}V_X^T,
\end{eqnarray}
is the unique minimizer to problem \eqref{eq:lr:nuclear_norm_minization}, where $V_X$ and $V_A$ are calculated as
follows: Compute the skinny Singular Value Decomposition (SVD) of $[X,A]$, denoted as $[X,A] =U\Sigma{}V^T$, and
partition $V$ as $V=[V_X;V_A]$ such that $X=U\Sigma{}V_X^T$ and $A=U\Sigma{V_A}^T$.
\end{theorem}
From the above theorem we have the following two corollaries. First, when the matrix $A$ is of full row rank, the
closed-form solution defined by \eqref{eq:lrr:close_form} can be represented in a simpler form.
\begin{corollary}\label{coro:unique_closed_form}
Suppose the matrix $A$ has full row rank, then
\begin{eqnarray*}
Z^*= A^T(AA^T)^{-1}X,
\end{eqnarray*}
is the unique minimizer to problem \eqref{eq:lr:nuclear_norm_minization}, where $A^T(AA^T)^{-1}$ is the
\emph{generalized inverse} of $A$.
\end{corollary}
Second, when the data matrix itself is used as the dictionary, i.e., $A=X$, the solution to problem
\eqref{eq:lr:nuclear_norm_minization} falls back to the outputs of a factorization based method.
\begin{corollary}\label{coro:A=X}
Assume $X\neq0$. Then the following optimization problem
\begin{eqnarray*}
\min_{Z} \hspace{0.1in}  \norm{Z}_*, &\text{s.t.}&X=XZ,
\end{eqnarray*}
has a unique minimizer
\begin{eqnarray}\label{eq:coro:close_form:sim}
Z^* = \mathrm{SIM}(X),
\end{eqnarray}
where $\mathrm{SIM}(X)=V_XV_X^T$ is called the Shape Interaction Matrix (SIM) \cite{ijcv_1998_factor} in computer
vision and $X=U_X\Sigma_XV_X^T$ is the skinny SVD of $X$.
\end{corollary}
\vspace{0.1cm}

The proof of Theorem \ref{theorem:unique:nonoise} is based on the following three lemmas.
\begin{lemma}\label{lemma:basic}
Let $U$, $V$ and $M$ be matrices of compatible dimensions. Suppose both $U$ and $V$ have orthogonal columns, i.e.,
$U^TU=\Id$ and $V^TV=\Id$ \footnote{Note here that $U$ and $V$ may not be orthogonal, namely, $UU^T\neq\Id$ and
$VV^T\neq\Id$.}, then we have
\begin{eqnarray*}
\norm{M}_* = \|UMV^T\|_*.
\end{eqnarray*}
\begin{proof}Let the full SVD of $M$ be $M=U_M\Sigma_MV_M^T$, then $UMV^T=(UU_M)\Sigma_{M}(VV_M)^T$.
As $(UU_M)^T(UU_M)=\Id$ and $(VV_M)^T(VV_M)=\Id$, $(UU_M)\Sigma_{M}(V_MV)^T$ is actually the SVD of $UMV^T$. By the
definition of the nuclear norm, we have $\norm{M}_* = \trace{\Sigma_M}=\norm{UMV^T}_*$.
\end{proof}
\end{lemma}

\begin{lemma}\label{lemma:abcd:0condition}
For any four matrices $B$, $C$, $D$ and $F$ of compatible dimensions, we have
\begin{eqnarray*}
\norm{\left[\begin{array}{cc}
B&C\\
D&F\\
\end{array}\right]}_* \geq \norm{B}_*,
\end{eqnarray*}
where the equality holds if and only if $C=0,D=0$ and $F=0.$
\end{lemma}
\begin{proof} Lemma 10 of \cite{lrr_psd} directly leads to the above conclusion.
\end{proof}

\begin{lemma}\label{lemma:unique:nonoise}
Let $U$, $V$ and $M$ be given matrices of compatible dimensions. Suppose both $U$ and $V$ have orthogonal columns,
i.e., $U^TU=\Id$ and $V^TV=\Id$, then the following optimization problem
\begin{eqnarray}\label{eq:lemma:problem}
\min_{Z} \hspace{0.1in} \norm{Z}_*, & \textrm{s.t.} & U^TZV=M,
\end{eqnarray}
has a unique minimizer
\begin{eqnarray*}
Z^* = UMV^T.
\end{eqnarray*}
\end{lemma}
\begin{proof}
First, we prove that $\norm{M}_*$ is the minimum objective function value and $Z^*=UMV^T$ is a minimizer. For any
feasible solution $Z$, let $Z = U_Z\Sigma_ZV_Z^T$ be its full SVD. Let $B=U^TU_Z$ and $C=V_Z^TV$. Then the constraint
$U^TZV=M$ is equal to
\begin{equation}\label{eq:lemma:BSzC=M}
B\Sigma_ZC = M.
\end{equation}
Since $BB^T=\Id$ and $C^TC=\Id$, we can find the orthogonal complements \footnote{When $B$ and/or $C$ are already
orthogonal matrices, i.e., $B_{\bot}=\emptyset$ and/or $C_{\bot}=\emptyset$, our proof is still valid.} $B_{\bot}$ and
$C_{\bot}$ such that
\begin{eqnarray*}
\left[\begin{array}{c}
B\\
B_{\bot}\\
\end{array}\right] &\textrm{and}&[C,C_{\bot}]
\end{eqnarray*}
are orthogonal matrices. According to the unitary invariance of the nuclear norm, Lemma \ref{lemma:abcd:0condition} and
\eqref{eq:lemma:BSzC=M}, we have
\begin{eqnarray*}
\norm{Z}_* &=& \norm{\Sigma_Z}_* = \norm{\left[\begin{array}{c}
B\\
B_{\bot}\\
\end{array}\right]\Sigma_Z[C,C_{\bot}]}_*=\norm{\left[\begin{array}{cc}
B\Sigma_ZC&B\Sigma_ZC_{\bot}\\
B_{\bot}\Sigma_ZC&B_{\bot}\Sigma_ZC_{\bot}\\
\end{array}\right]}_*\\
&\geq&\norm{B\Sigma_ZC}_*=\norm{M}_*,
\end{eqnarray*}
Hence, $\norm{M}_*$ is the minimum objective function value of problem \eqref{eq:lemma:problem}. At the same time,
Lemma \ref{lemma:basic} proves that $\norm{Z^*}_*=\norm{UMV^T}_*=\norm{M}_*$. So $Z^*=UMV^T$ is a minimizer to problem
\eqref{eq:lemma:problem}.

Second, we prove that $Z^*=UMV^T$ is the unique minimizer. Assume that $Z_1=UMV^T+H$ is another optimal solution. By
$U^TZ_1V=M$, we have
\begin{equation}\label{eq:lemma:UHV=0}
U^THV = 0.
\end{equation}
Since $U^TU=\Id$ and $V^TV=\Id$, similar to above, we can construct two orthogonal matrices: $[U,U_{\bot}]$ and
$[V,V_{\bot}]$. By the optimality of $Z_1$, we have
\begin{eqnarray*}
\norm{M}_*&=&\norm{Z_1}_*=\norm{UMV^T+H}_*=\norm{\left[\begin{array}{c}
U^T\\
U_{\bot}^T\\
\end{array}\right](UMV^T+H)[V,V_{\bot}]}_*\\
&=&\norm{\left[\begin{array}{cc}
M&U^THV_{\bot}\\
U_{\bot}^THV&U_{\bot}^THV_{\bot}\\
\end{array}\right]}_*\geq\norm{M}_*.
\end{eqnarray*}
According to Lemma \ref{lemma:abcd:0condition}, the above equality can hold if and only if
\begin{equation*}
U^THV_{\bot}=U_{\bot}^THV=U_{\bot}^THV_{\bot}=0.
\end{equation*}
Together with \eqref{eq:lemma:UHV=0}, we conclude that $H=0$. So the optimal solution is unique.
\end{proof}
\vspace{0.1cm}

The above lemma allows us to get closed-form solutions to a class of nuclear norm minimization problems. This leads to
a simple proof of Theorem \ref{theorem:unique:nonoise}. \vspace{0.1cm}

\begin{proof}{(\emph{of Theorem \ref{theorem:unique:nonoise}})}
Since $X\in{\spn{A}}$, we have $\rank{[X,A]}=\rank{A}$. By the definitions of $V_X$ and $V_A$ (see Theorem
\ref{theorem:unique:nonoise}), it can be concluded that the matrix $V_A^T$ has full row rank. That is, if the skinny
SVD of $V_A^T$ is $U_1\Sigma_1V_1^T$, then $U_1$ is an orthogonal matrix. Through some simple computations, we have
\begin{equation}\label{eq:theorem:inv_va}
V_A(V_A^TV_A)^{-1} = V_1\Sigma_1^{-1}U_1^T.
\end{equation}
Also, it can be calculated that the constraint $X=AZ$ is equal to $V_X^T=V_A^TZ$, which is also equal to
$\Sigma_1^{-1}U_1^TV_X^T=V_1^TZ$. So problem \eqref{eq:lr:nuclear_norm_minization} is equal to the following
optimization problem:
\begin{eqnarray*}
\min_{Z}\hspace{0.1in}  \norm{Z}_*, & \textrm{s.t.} & V_1^TZ=\Sigma_1^{-1}U_1^TV_X^T.
\end{eqnarray*}
By Lemma \ref{lemma:unique:nonoise} and \eqref{eq:theorem:inv_va}, problem \eqref{eq:lr:nuclear_norm_minization} has a
unique minimizer
\begin{eqnarray*}
Z^* &=& V_1\Sigma_1^{-1}U_1^TV_X^T=V_A(V_A^TV_A)^{-1}V_X^T.
\end{eqnarray*}
\end{proof}
\section{Conclusion}In this paper, we prove that problem \eqref{eq:lr:nuclear_norm_minization} has a unique and
closed-form solution. The heart of the proof is Lemma \ref{lemma:unique:nonoise}, which actually allows us to get the
closed-form solutions to a category of nuclear norm minimization problems. For example, by following the clues
presented in this paper, it is simple for one to get the closed-form solution to the following optimization problem:
\begin{eqnarray*}
\min_{Z}\hspace{0.1in}  \norm{Z}_*, & \textrm{s.t.} & X=AZB,
\end{eqnarray*}
where $X$, $A$ and $B$ are given matrices. Our theorems are useful for studying the LRR problems. For example, based on
Lemma \ref{lemma:unique:nonoise}, we have devised an method to recovery the effects of the unobserved, hidden data in
LRR \cite{latlrr_cvpr_2010}.
\bibliographystyle{IEEEtran}
\bibliography{latlrr}

\end{document}